\documentclass[11pt]{article}

\usepackage[margin=1in]{geometry}
\usepackage{amsmath,amssymb}
\usepackage{booktabs}
\usepackage{graphicx}
\usepackage{microtype}
\usepackage{url}
\usepackage[hidelinks]{hyperref}

\title{Reliable Federated TinyML Deployment for IoT Security}

\author{%
Younsoo Park$^{1}$ \quad Seokhyoen Bae$^{1}$ \quad Shasi Kumar Ramachandran Prabhu$^{1}$\\
Suman Saha$^{1}$ \quad Peilong Li$^{2}$\\[0.75em]
\small $^{1}$The Pennsylvania State University, University Park, Pennsylvania, USA\\
\small $^{2}$Elizabethtown College, Elizabethtown, Pennsylvania, USA\\[0.35em]
\small \texttt{danielpark0605@gmail.com}, \texttt{baeandew@gmail.com}, \texttt{sfr5846@psu.edu},\\
\small \texttt{szs339@psu.edu}, \texttt{lip@etown.edu}
}

\date{}

\begin{document}

\maketitle

\begin{abstract}
The growing deployment of Internet of Things (IoT) devices has increased the need for privacy-preserving intrusion detection systems that operate directly on resource-constrained hardware. Federated Learning enables collaborative model training without sharing raw data, but conventional federated models are often too large and unstable for deployment on microcontroller-class devices. TinyML techniques enable compact neural networks but are typically designed for inference-only workloads.

This work investigates combining Federated Learning with TinyML-based model compression for intrusion detection in IoT environments. We evaluate compression strategies including knowledge distillation, structured pruning, and quantization within a federated training pipeline. Preliminary results show that training stability plays a critical role in federated TinyML systems. In particular, server-coordinated cosine learning-rate scheduling improves Attack Recall from 46.7\% to 93.85\% while enabling substantial model compression and efficient edge deployment. These findings provide insights for designing lightweight and privacy preserving intrusion detection systems for IoT devices.
\end{abstract}

\noindent\textbf{Keywords:} Federated Learning, TinyML, Intrusion Detection Systems, IoT Security, Model Compression

\section{Introduction}

The rapid proliferation of Internet of Things (IoT) devices has enabled large-scale deployment of data-driven applications across domains such as smart homes, healthcare, and industrial systems \cite{atzori2010iot}. These applications increasingly rely on machine learning models trained on sensitive data, raising concerns about privacy, data ownership, and communication overhead in centralized training.

Federated Learning (FL) addresses these challenges by enabling collaborative training without sharing raw data \cite{mcmahan2017communication, kairouz2021advances, li2020fedprox}. However, deploying FL in IoT environments remains difficult due to limited memory, computation, and energy resources, as well as the large model sizes typically assumed in standard FL pipelines \cite{konevcny2016federated}.

TinyML enables neural networks to run directly on microcontrollers by optimizing model size, memory footprint, and inference latency \cite{banbury2020tinyml}. However, most TinyML systems focus on inference-only workloads, and integrating TinyML with federated training remains an open challenge, particularly for security-critical applications such as intrusion detection.

Machine learning-based intrusion detection systems (IDS) can achieve strong detection performance \cite{javaid2016deep}, but deploying them on IoT devices requires lightweight models while maintaining high attack recall. In our early experiments on the CIC-IDS2017 dataset, a federated IDS model trained with a fixed learning rate achieved only 46.7\% attack recall, despite high overall accuracy, due to unstable optimization under class imbalance. Additional challenges include deployment instability (e.g., BatchNorm interactions) and performance degradation under aggressive compression \cite{han2016deepcompression,hinton2015distilling,jacob2018quantization}.

To address these challenges, we investigate combining FL with TinyML-oriented compression techniques for IoT intrusion detection. We evaluate knowledge distillation, structured pruning, quantization-aware training (QAT), and post-training quantization (PTQ), and analyze their impact on detection performance, training stability, and deployment feasibility. We further examine adversarial robustness using the Fast Gradient Sign Method (FGSM) \cite{goodfellow2015fgsm}.

Our results show that training stability is a key factor in federated TinyML systems. In particular, server-coordinated cosine learning-rate scheduling improves attack recall from 46.7\% to 93.85\%. In addition, the proposed compression pipeline significantly reduces model size and inference latency while preserving strong detection performance. These findings provide practical guidelines for designing lightweight and privacy-preserving intrusion detection systems for resource-constrained IoT devices.

\section{Related Work}

Federated Learning (FL) enables collaborative model training across distributed clients without sharing raw data \cite{mcmahan2017communication, kairouz2021advances}. This paradigm is well suited for IoT environments with privacy and bandwidth constraints, but practical deployment remains challenging due to communication overhead, client heterogeneity, and the difficulty of running large models on resource-constrained devices \cite{konevcny2016federated}.

TinyML addresses these limitations by enabling neural networks to run directly on microcontrollers through model size, memory, and latency optimization \cite{banbury2020tinyml}. However, most TinyML systems focus on inference-only workloads, and integrating TinyML with federated training requires lightweight models and carefully designed pipelines.

Model compression techniques are widely used to enable efficient deployment on constrained hardware. Structured pruning removes redundant parameters, knowledge distillation transfers knowledge from large teacher models to compact students, and quantization reduces numerical precision for efficient inference \cite{han2016deepcompression,hinton2015distilling,jacob2018quantization}. Quantization-aware training (QAT) and post-training quantization (PTQ) are commonly used to convert floating-point models into efficient integer representations.

Machine learning-based intrusion detection systems (IDS) have shown strong performance in detecting network attacks, but most approaches assume centralized training and resource-rich environments. In addition, robustness to adversarial attacks has been widely studied, with methods such as the Fast Gradient Sign Method (FGSM) and Projected Gradient Descent (PGD) commonly used for evaluation \cite{goodfellow2015fgsm}.

Despite these advances, few studies have jointly addressed federated learning, TinyML deployment constraints, and intrusion detection in IoT systems. Achieving high detection performance while maintaining compact models and reliable embedded deployment remains an open challenge. This work addresses this gap by evaluating compression strategies and training techniques within a federated TinyML pipeline for resource-constrained IoT intrusion detection.
\section{Approach}

The goal of this work is to maintain high intrusion detection performance while minimizing model size to enable deployment on resource-constrained IoT devices, while preserving accuracy and F1-score.

\subsection{Federated Training Baseline}

We begin with a baseline Federated Learning (FL) framework in which edge devices locally train an intrusion detection model and periodically transmit model updates to a central server, which aggregates updates to redistributed to clients \cite{mcmahan2017communication, kairouz2021advances, li2020fedprox}. To address class imbalance in intrusion detection data, we employ Focal Loss

\[
FL(p_t)=-\alpha_t(1-p_t)^\gamma\log(p_t),
\]

where $p_t$ is the predicted probability for the ground-truth class. In our experiments we set $\alpha=0.7$ and $\gamma=2.0$. Training runs for 80 federated communication rounds and establishes the reference performance before compression. Early experiments revealed unstable convergence under a fixed learning rate due to dataset imbalance. To stabilize training we introduce server-coordinated cosine learning-rate decay \cite{loshchilov2017sgdr}, where the global learning rate $\eta$ evolves across $T$ rounds as

\[
\eta_t=\eta_{\min}+\frac{1}{2}(\eta_{\max}-\eta_{\min})(1+\cos(t\pi/T)).
\]

\subsection{Model Compression Experiments}

To reduce model size while maintaining detection performance, we evaluate compression techniques including knowledge distillation and structured pruning \cite{hinton2015distilling,han2016deepcompression}. Multiple configurations are explored to identify the best trade-off between detection accuracy and model efficiency.

The final compression pipeline consists of BatchNorm folding, knowledge distillation, structured pruning, and optional quantization-aware fine-tuning before conversion to TFLite INT8 format. Each configuration is evaluated using model size, accuracy, and F1-score to determine which compression strategies best support deployment on resource-constrained IoT hardware.

\subsection{Model Selection for Deployment}

From the compression experiments we select four representative configurations illustrating different performance–efficiency trade-offs:

\begin{itemize}
\item Most Compressed: smallest model with acceptable performance
\item Most Accurate: highest accuracy and F1-score
\item Balanced-Compressed: moderate compression with strong performance
\item Balanced-Accurate: high accuracy with moderate model size
\end{itemize}

These configurations allow practitioners to select models based on deployment constraints. Highly constrained devices may require the most compressed configuration, while more capable devices can adopt balanced or accuracy-oriented models.
\section{Experimental Results}

We report experimental results for federated TinyML intrusion detection, focusing on training stability, compression efficiency, quantization strategies, and deployment reliability. All experiments use the CIC-IDS2017 dataset with four federated clients and focal loss to address class imbalance.

\subsection{Experimental Setup}

All experiments use the CIC-IDS2017 dataset\cite{sharafaldin2018cicids}\footnote{Canadian Institute for Cybersecurity Intrusion Detection System Dataset (CIC-IDS2017): \url{https://www.unb.ca/cic/datasets/ids-2017.html}}. Alternative datasets such as Bot-IoT and TON\_IoT were considered but excluded due to extreme class imbalance. The dataset is split into an 80:20 train/test partition using stratified sampling. Majority-class undersampling is applied to the training data to obtain an approximate 80:20 normal-to-attack ratio, while the test set remains unchanged. We simulate a federated learning environment with four clients following the standard FL paradigm \cite{mcmahan2017communication}. Each client trains locally while the server aggregates updates using FedAvgM and applies cosine learning-rate decay to stabilize training.

\subsection{Compression Pipeline}

The compressed models are produced using a multi-stage compression pipeline. As summarized in Table~\ref{tab:compression}, the final compressed model reduces model size from 0.78 MB to 0.0635 MB (12.28$\times$ compression) and lowers inference latency from 1.89 ms to 0.48 ms, corresponding to a 74.5\% latency reduction. 

First, BatchNorm folding merges BatchNormalization parameters into the preceding Dense layer. The updated weights $W_{fold}$ and bias $b_{fold}$ are computed from $(\gamma,\beta,\mu,\sigma)$ as

\[
W_{fold}=\frac{\gamma W}{\sqrt{\sigma^2+\epsilon}}, \qquad
b_{fold}=\beta+\frac{\gamma (b-\mu)}{\sqrt{\sigma^2+\epsilon}}.
\]

Next, a smaller student model is trained using progressive knowledge distillation, where the federated model serves as the teacher. The student network contains 50\% of the original parameters. Structured pruning is then applied to remove low-importance neurons from Dense layers. Finally, a quantization-aware fine-tuning step is optionally applied to prepare the model for INT8 deployment.

\subsection{Training Stability and Detection Performance}
\label{sec:training-stability}

Early experiments on the CIC-IDS2017 dataset revealed that a naive federated training setup with a fixed learning rate leads to highly unstable optimization. Although the global model often achieved high overall accuracy, its recall for the attack class remained low: in a representative run, the model correctly detected only 46.7\% of attack flows, indicating that many malicious connections were systematically misclassified as benign. To address this issue, we combine focal loss with a server-coordinated cosine learning-rate schedule. Focal loss down-weights easy benign samples and amplifies gradients from hard attack examples, while cosine decay anneals the global learning rate over $T$ federated communication rounds to avoid late-stage oscillations. With this configuration, our federated baseline improves recall for the attack class from 46.7\% to 93.85\% on CIC-IDS2017, while maintaining high overall accuracy and eliminating divergence observed in the fixed learning-rate runs.

This stable training regime serves as the foundation for our compression study. Starting from the converged federated model, we apply our compression pipeline to generate 48 configurations that combine FL QAT, knowledge distillation (none, direct, progressive), structured pruning (\texttt{none}, \texttt{10x5}, \texttt{10x2}, \texttt{5x10}), and optional post-training quantization. Across these compressed models, most configurations preserve recall close to the 93.85\% baseline before compression. In particular, well-performing configurations retain recall above 90\% while significantly reducing model size. This indicates that our compression pipeline enables substantial model size reduction while preserving detection performance.

\begin{table}[t]
\centering
\caption{Detection performance before and after compression.}
\label{tab:performance}
\begin{tabular}{lccc}
\toprule
Model & Accuracy & F1-score & Recall (Attack) \\
\midrule
Baseline Model & 93.5\% & 84.1\% & 46.7\% \\
Compressed Model & \textbf{96.02\%} & \textbf{89.32\%} & \textbf{93.85\%} \\
\bottomrule
\end{tabular}
\end{table}

\begin{table}[t]
\centering
\caption{Detection performance and corresponding compression configurations under moderate and extreme compression.}
\label{tab:compression}
\resizebox{\textwidth}{!}{%
\begin{tabular}{lcccccccc}
\toprule
Compression & Method & FL QAT & Distillation & Pruning & PTQ & Model Size & Accuracy & F1-score \\
\midrule
Moderate & No QAT during FL & No & Progressive & $10\times5$ & No & 37.89 KB & \textbf{99.16\%} & \textbf{97.60\%} \\
Moderate & QAT during FL & Yes & Progressive & $10\times5$ & No & 37.91 KB & 97.91\% & 94.33\% \\
Extreme & No QAT during FL & No & Progressive & $5\times10$ & Yes & 14.44 KB & 93.54\% & 77.63\% \\
Extreme & QAT during FL & Yes & Progressive & $5\times10$ & Yes & 14.44 KB & \textbf{95.81\%} & \textbf{86.66\%} \\
\bottomrule
\end{tabular}%
}
\end{table}

\subsection{Detection Performance}

Table~\ref{tab:performance} summarizes detection performance before and after applying the compression pipeline. The baseline federated model exhibits unstable convergence and low recall for the attack class (46.7\%). After introducing server-coordinated cosine learning-rate decay, detection performance improves substantially, reaching 96.02\% accuracy, 89.32\% F1-score, and 93.85\% recall for the attack class.

The compressed model is obtained by applying BatchNorm folding, knowledge distillation, structured pruning, and quantization-aware fine-tuning to this stabilized model. Therefore, the performance improvement primarily results from the stabilized training process, while the compression pipeline reduces model size and latency while preserving detection performance.

\subsection{Compression and Quantization Analysis}

We analyze the interaction between quantization, knowledge distillation, and structured pruning in the federated TinyML compression pipeline. In total, 48 configurations are evaluated, comparing two strategies: training with and without quantization-aware training (QAT) during federated learning. Table~\ref{tab:compression} summarizes compression configurations and detection performance under moderate and extreme compression levels. Under moderate compression (approximately 38\,KB), the configuration without QAT achieves the best performance, reaching 99.16\% accuracy and 97.60\% F1-score. Enabling QAT during federated training slightly reduces detection performance at this compression level. However, when compression is pushed further to approximately 14\,KB models, the trend reverses. In this regime, the configuration using QAT achieves higher accuracy (95.81\%) and F1-score (86.66\%) compared to the no-QAT configuration. These results suggest that the benefit of QAT depends on the compression regime. While full-precision training performs better under moderate compression, QAT becomes important under extreme compression, where it helps the model adapt to quantization noise and maintain detection performance.

These observations suggest that quantization-aware training (QAT) is not always beneficial during the federated training stage. In our experiments, configurations trained without QAT during FL sometimes achieve higher detection performance under moderate compression levels. However, when models are aggressively compressed, QAT helps stabilize the network against quantization noise and improves performance under aggressive compression.

This behavior is consistent with prior studies showing that QAT may introduce optimization instability due to oscillations between quantization grid points during training \cite{nagel2022overcoming}. Moreover, several analyses indicate that post-training quantization (PTQ) can achieve near floating-point performance at 8-bit precision without requiring QAT in many practical scenarios \cite{nagel2021white}. These findings highlight that the benefit of QAT depends strongly on the compression regime and training configuration.
\section{Future Work}

Future work will focus on three directions.

\noindent \textbf{\textit{Dynamic Communication and Adaptive Aggregation:}} We plan to reduce communication overhead by exploring selective client participation, where only devices with significant updates transmit model parameters. We will also investigate asynchronous federated aggregation to handle heterogeneous network conditions across IoT nodes.

\noindent \textbf{\textit{Adversarial Robustness and Training Strategies:}} While this work focuses on improving training stability and compression efficiency, future work will extend the framework to include adversarial robustness. In particular, we plan to incorporate adversarial training techniques within the federated learning pipeline and evaluate robustness under stronger attacks such as Projected Gradient Descent (PGD). This will allow us to study the trade-off between model compression, detection performance, and robustness in resource-constrained environments.

\noindent \textbf{\textit{Real-World Deployment on Microcontrollers:}} We will further validate the proposed pipeline through end-to-end deployment on microcontroller-class devices such as ESP32 using TensorFlow Lite for Microcontrollers. This includes measuring on-device inference latency, memory usage, and energy consumption to ensure that the compressed models operate reliably under real-world hardware constraints.
\section{Conclusion}

This work investigates the feasibility of deploying federated intrusion detection models on resource-constrained IoT devices by integrating TinyML-oriented compression techniques. Our results show that stabilizing federated training plays a critical role in improving detection performance, increasing Attack Recall from 46.7\% to 93.85\% using server-coordinated cosine learning-rate scheduling. The proposed compression pipeline further enables efficient edge deployment, achieving a 12.28$\times$ reduction in model size and a 74.5\% reduction in inference latency on ESP32-class hardware while preserving detection accuracy. 

Our analysis also reveals that the effectiveness of quantization-aware training depends on the compression regime: full-precision training performs better under moderate compression, whereas QAT becomes important when models are aggressively compressed. These findings provide practical guidance for designing lightweight and privacy-preserving intrusion detection systems suitable for federated TinyML deployment in IoT environments.

\bibliographystyle{unsrt}
\bibliography{references}

\end{document}